\documentclass[conference]{IEEEtran}
\IEEEoverridecommandlockouts
\usepackage{cite}
\usepackage{amsmath,amssymb,amsfonts}
\usepackage{algorithmic}
\usepackage{graphicx}
\usepackage{textcomp}
\usepackage{xcolor}
\def\BibTeX{{\rm B\kern-.05em{\sc i\kern-.025em b}\kern-.08em
    T\kern-.1667em\lower.7ex\hbox{E}\kern-.125emX}}
\begin{document}

\def\@IEEEfigurecaptionsepspace{\vskip 0pt}
\setlength{\textfloatsep}{4pt}   % Space between float e texto
\setlength{\abovecaptionskip}{-4pt}   % Reduz o espaçamento entre a figura e a legenda
\setlength{\floatsep}{4pt}       % Space between floats consecutivos

\title{Cross-Platform Neural Video Coding: A Case Study}
% \title{Hyperprior Decoding Path Quantization in SSF Codec for Mismatch-Aware End-to-End Compression}

\author{\IEEEauthorblockN{Ruhan Conceição\IEEEauthorrefmark{2}\IEEEauthorrefmark{3}, Marcelo Porto\IEEEauthorrefmark{2}, Wen-Hsiao Peng\IEEEauthorrefmark{3}, and Luciano Agostini\IEEEauthorrefmark{2}}
\IEEEauthorblockA{\IEEEauthorrefmark{2}Graduate Program in Computing, Federal University of Pelotas\\
\IEEEauthorrefmark{3}Departament of Computer Science, National Yang Ming Chiao Tung University\\
radconceicao@inf.ufpel.edu.br, porto@inf.ufpel.edu.br, wpeng@cs.nctu.edu.tw, agostini@inf.ufpel.edu.br} \vspace{-2em} 
% \and
% \IEEEauthorblockN{Wen-Hsiao Peng}
% \IEEEauthorblockA{\textit{College of Computer Science} \\
% \textit{National Yang Ming Chiao Tung University}\\
% Hsinchu, Taiwan \\
% wpeng@cs.nctu.edu.tw}
% \and
% \IEEEauthorblockN{Luciano Agostini}
% \IEEEauthorblockA{\textit{dept. name of organization (of Aff.)} \\
% name of organization (of Aff.)\\
% Pelotas/RS, Brazil \\
% agostini@inf.ufpel.edu.br}
}

\maketitle

\begin{abstract}
In this paper, we first show that current learning-based video codecs, specifically the SSF codec, are not suitable for real-world applications due to the mismatch between the encoder and decoder caused by floating-point round-off errors. To address this issue, we propose the static quantization of the hyper prior decoding path. The quantization parameters are determined through an exhaustive search of all possible combinations of observers and quantization schemes from PyTorch. For the SSF codec, when encoding and decoding on different machines, the proposed solution effectively mitigates the mismatch issue and enhances compression efficiency results by preventing severe image quality degradation. When encoding and decoding are performed on the same machine, it constrains the average BD-rate increase to 9.93\% and 9.02\% for UVG and HEVC-B sequences, respectively.

\end{abstract}


\section{Introduction}
Learning-based video codecs, while offering significant advancements in compression efficiency, still encounter several challenges, such as high computational cost and extensive memory access, that hamper their deployment in real-world applications. 
%One of the primary concerns is the high computational cost required to both encode and decode video sequences~\cite{ComplexityNVC}. Moreover, certain codecs demand extensive memory access to read and write feature maps, which can become prohibitive due to hardware limitations and increased power consumption. 
Among these challenges, cross-platform inconsistencies in floating-point arithmetic -- the focus of this work -- present a significant obstacle, complicating the adoption of these codecs across diverse devices and environments.

Since the release of the Deep Video Compression (DVC)\cite{DVC}, several neural network-based autoencoders for video compression have been proposed in the literature. The Scale-Space Flow (SSF)\cite{SSF} improves the DVC by generalizing the optical flow with the addition of a scale parameter, allowing the network to better handle varying levels of motion uncertainty and thereby increasing compression efficiency. Instead of employing conventional residual coding, the Deep Contextual Video Coding (DCVC)~\cite{DCVC} implements conditional coding. This codec has several extensions like \cite{DCVC-TCM,DCVC-HEM,DCVC-DC}, to learn more effectively temporal contextual information and enhance entropy coding. More recently, the Masked Conditional Residual Transformer (MaskCRT)\cite{MaskCRT} was proposed, adopting a new coding framework known as conditional residual coding. In common, these learned video codecs feature a variational autoencoder~\cite{VAE} (VAE)-based compression backbone along with a hyper prior~\cite{Balle_VAE_Hyper}, which models the probability distributions of the main latents for their entropy coding.

\begin{figure*}[tbp]
    \centerline{\includegraphics[width=.90\textwidth]{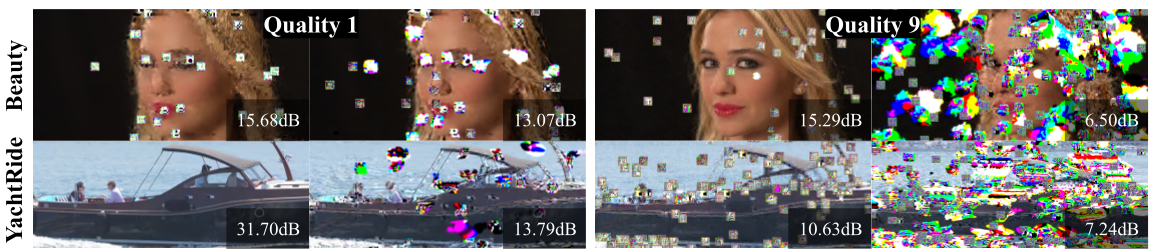}}
    \caption{Floating-point error propagation across different SSF quality levels. Full-resolution images are available at https://ssfquant-iscas2025.netlify.app/}
    \label{fig_fp_error}
\end{figure*}

Both the encoder and decoder must evaluate the prior distributions over the main latents and must have identical results~\cite{Balle_Integer}. Small perturbations caused by floating-point round-off errors can lead to significant error propagation, causing catastrophic degradation in decoded video quality (see Fig.~\ref{fig_fp_error}). However, learning-based video codecs often rely on floating-point arithmetic, which is strongly platform-dependent. How to ensure fully synchronized encoding and decoding results when encoding and decoding are performed on different platforms remains largely under-explored for learned video codecs.

In this context, we present an evaluation of the impact of floating-point round-off errors on a typical learned video codec, the SSF~\cite{SSF}, by encoding and decoding video sequences on different machines and operating systems. Additionally, this work evaluates the effect of quantizing the hyper prior coding path using an 8-bit integer representation, analyzing its ability to suppress representation errors and and its impact on coding efficiency. To the best of our knowledge, this is one of the few early attempts at analyzing and addressing floating-point round-off errors for end-to-end learned video codecs although they have been similar studies for learned image codecs.

\section{Background}

\subsection{The SSF Codec} \label{sec_ssf_cai}
\begin{figure}[tbp]
    \centering
    \includegraphics[width=0.95\linewidth]{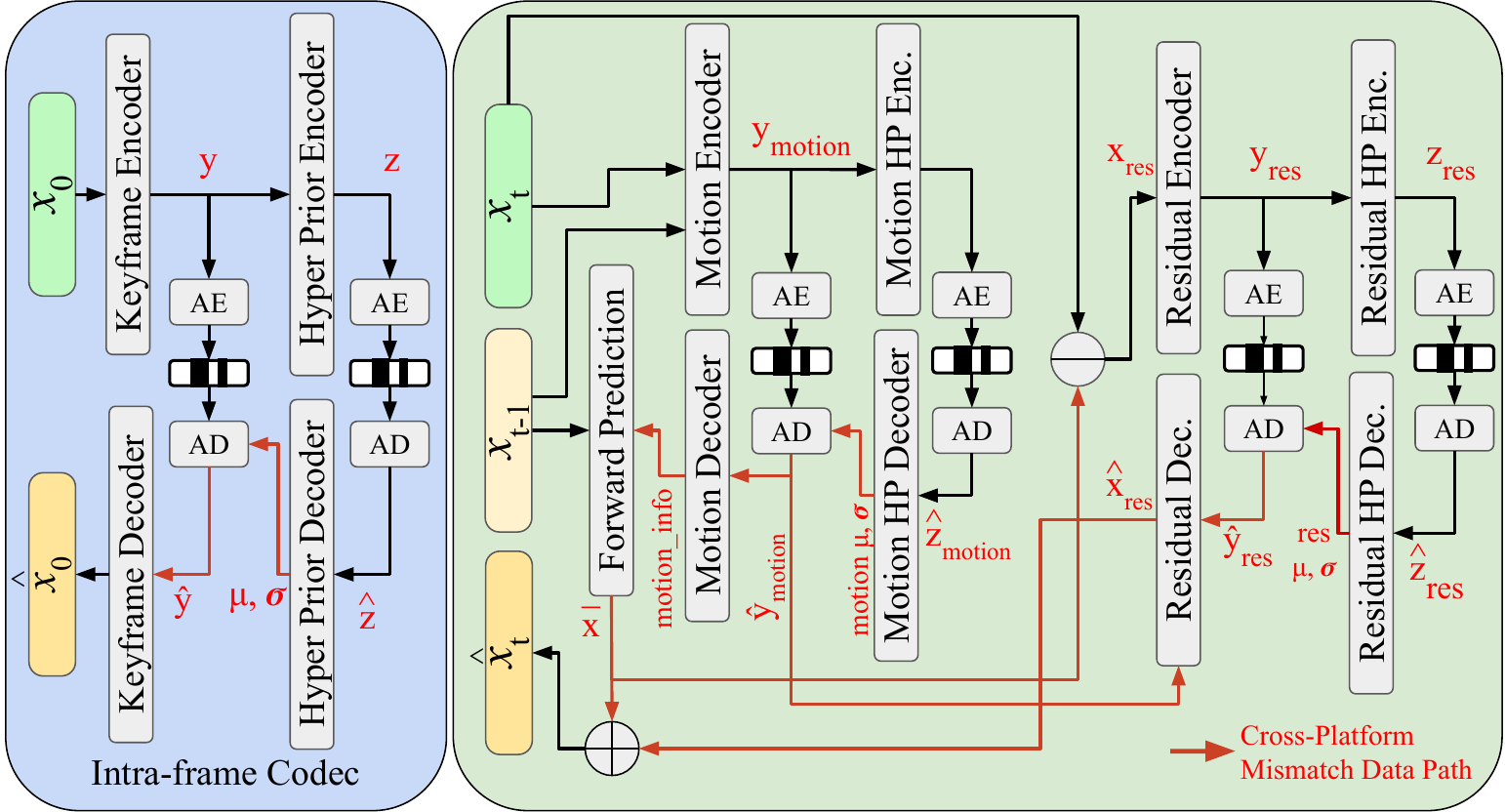}
    \caption{SSF Codec Block Diagram}
    \label{fig_ssf}
\end{figure}

The Scale-Space Flow (SSF)\cite{SSF} is a representative learning-based video codec. It was designed to improve both motion estimation and compensation. The SSF generalizes optical flow by incorporating a scale parameter, which allows the network to model motion uncertainty across spatial locations. 
%This approach enables the network to model a wide variety of motion within the video. 
After estimating the motion, the network warps the reference frame to match the current frame, minimizing the inter-frame redundancy. As a residual codec, the SSF encodes the difference between the predicted and current frame through a neural network, and transmits it in the encoded bitstream. The SSF incorporates a hyper prior model. Fig.~\ref{fig_ssf} presents a block diagram of the SSF. The figure also labels the intermediate and latent codes. In this context, AE and AD represent the arithmetic encoder and decoder, respectively, while HP stands for hyper prior. The SSF from 
CompressAI~\cite{CompressAI} currently supports nine quality levels, each of which is produced by a separate model.
%is an open-source library that provides pre-trained models and tools for evaluating and implementing learning-based image and video compression algorithms. In CompressAI, SSF 
%, illustrating both the intra-frame and inter-frame codecs and their main latent and hyper prior paths

% Recently, some works in the literature was published aiming at addressing this issue. The authors of~\cite{Towards_RealTime} proposed cross-platform video codec system which sends calibration side information to ensure the consistent quantization of entropy parameters. In~\cite{Towards_Reproducible}, the authors proposes a safeguard mechanism for image and point-clouds that identifies and protect variables near to critical points in the quantization stage.

\subsection{Literature Review on Cross-platform Mismatch Mitigation}
Recent research on addressing the issue of floating-point round-off errors in neural networks has been focused on two main approaches: quantization and extra signaling.

\subsubsection{Quantization}
Ballé \textit{et al}.~\cite{Balle_Integer} proposed an integer neural network for learned image compression to mitigate cross-platform floating-point round-off errors. They implemented all neural network layers with integer arithmetic, restricting operations to basic arithmetic or lookup tables to ensure consistency across platforms. Sun \textit{et al}.\cite{HSun_Integer} introduced an end-to-end image compression method using fixed-point weights and activations. They explored various quantization schemes and identified a channel-wise non-linear quantization approach based on a coding gain analysis. To further reduce quantization errors, the authors proposed multiple non-linear quantization codebooks with varying dynamic ranges for different activation ranges. In another study\cite{HSun_QLIC}, Sun \textit{et al.} introduced the channel splitting technique, which reduces quantization errors by splitting channels with a high reconstruction error impact, allowing for more efficient quantization.

\subsubsection{Extra Signaling} In~\cite{Towards_Reproducible}, the authors introduced a ``safeguarding bitstream'' approach, where the encoder signals ``risky flags'' for variables near quantization boundaries that could cause discrepancies. The decoder uses these flags to adjust the variables and ensure that its latent code matches exactly what the encoder produces. Similarly, in~\cite{Towards_RealTime}, the authors introduced a calibration information transmission system to ensure cross-platform consistency in video codecs. By identifying the entropy parameters prone to errors due to floating-point discrepancies, the encoder transmits their coordinates in the bitstream, enabling the decoder to process these parameters accurately.

While these works propose relevant solutions to avoid representation mismatches and enable the cross-platform interoperability, they all are limited to learned image codecs, except~\cite{Towards_RealTime}. \cite{Towards_RealTime} requires the transmission of extra data and does not implement quantization. In contrast, the works proposed in~\cite{Balle_Integer} and~\cite{HSun_Integer} require quantization-aware training, meaning the neural networks must be trained from scratch to incorporate the proposed quantization schemes. In this work, we explore and implement a post-quantization scheme for video codecs that eliminates the need to retrain the encoder from scratch and avoid the transmission of extra information in the bitstream, requiring only a simple calibration process.

\subsection{PyTorch Built-in Quantization Framework} \label{sec_pytorch}

This work employs the built-in quantization framework provided by PyTorch library~\cite{Pytorch_Quant}. It supports several types of quantization, such as dynamic quantization and static quantization. We apply static quantization, which quantizes both weights and activations ahead of time. This method applies fixed quantization parameters to both weights and activations during inference. 

The first step in static quantization is to define where the data will be converted from the floating-point format to the 8-bit integer format, and vice versa. Then, the second step inserts observers into the model. These observers capture the distributions of weights and activations, which is followed by the third step: the calibration process. In the calibration step, representative data are passed through the model to collect statistics on these distributions. Once calibration is complete, the model is quantized by applying the parameters obtained from the observers.

Observers are employed to track the distributions of the activations and weights during the calibration process. They determine the scaling factors and zero points for converting floating-point values to integers and vice versa. PyTorch offers three types of observers: \textit{MinMax}, \textit{MovingAverageMinMax}, and \textit{Histogram}. The \textit{MinMax} observer records the minimum and maximum values of the activations or weights in the calibration stage. The \textit{MovingAverageMinMax} observer is a variant of the \textit{MinMax}. It computes a moving average of the minimum and maximum values over multiple batches that occur during the calibration step. Finally, the \textit{Histogram} observer uses a histogram to track the data distribution. It allows more fine-grained control of how values are distributed, focusing on the more frequently occurring values while ignoring outliers. 

Quantization schemes (QS) further determine how observed data are mapped to integer representations. PyTorch supports two main schemes: \textit{symmetric} and \textit{affine} (asymmetric). Symmetric quantization centers the integer range around zero, while affine quantization allows the zero point to be shifted to a non-centered position. Additionally, quantization can be applied \textit{per tensor} or \textit{per channel}. In the per-tensor scheme, a single scale and zero point are applied to the entire tensor, while in the per-channel scheme, each output channel has its own scale and zero point.

\section{Floating-point Round-off Errors} \label{sec_fp_error}

This section presents an analysis of floating-point round-off errors when running the encode-decode pipeline on distinct setups. Specifically, the encoding process was performed on an Intel\textregistered~Core™ i9-7920X CPU @ 2.90GHz running a Linux system whlist the decoding process was conducted on an Intel\textregistered~Core™ i5 CPU @ 1.80GHz running a MacOS system. This analysis was performed using the first 10 frames of the UVG dataset~\cite{UVG} and evaluated across all nine quality points of the non-quantized SSF pre-trained network. 

Fig.~\ref{fig_fp_error} shows the first and fifth frames of the decoded sequences \textit{Beauty} and \textit{YachtRide} at quality levels 1 and 9, in addition to their PSNR results. Note that no error is observed in the first frame of the \textit{YachtRide} sequence at quality level 1. However, once an error occurs, it propagates to next frames due to inter-frame prediction. This error propagation is noticeable in the fifth frame of the \textit{Beauty} sequence, where the error area seems to be smoother due to the use of motion compensation. Additionally, one may notice that the error becomes even more severe at quality level 9. By the fifth frame of both sequences, the scene becomes almost unidentifiable. It is worth mentioning that floating-point errors occurred across all evaluated sequences and quality levels. This fact demonstrates that the same issue highlighted by Ballé in~\cite{Balle_Integer} also occurs in learned video codecs, specifically in SSF. %It reinforces the need for investigations in the endeavor to mitigate the floating-point round-off errors in these scenarios, thus making it feasible to deploy these codecs in real-world environments where encoders and decoders operate on different platforms. 

%\subsection{Analysis of SSF Main Latents and Hyper Prior}
To identify where the mismatch in cross-platform processing occurs, we traced all signals shown in Fig.~\ref{fig_ssf} within the SSF decoder and compared them while decoding five frames of the \textit{Beauty} sequence on the same machine versus a different platform. The mismatches are highlighted in Fig.~\ref{fig_ssf} as red arrows. Through this analysis, we observed that all instances of $\hat{z}$ remained identical for intra-frame, motion, and residual decoding. The arithmetic decoder does not generate the cross-platform mismatch.
However, one may notice that both $\mu$ and $\sigma$ presented some discrepancies for intra-frame, motion, and residual decoding. This error then propagated through the other modules, as depicted. These findings support the hypothesis that the hyper prior decoder is likely the source of the error, preventing the decoder from producing consistent values across different platforms.

\section{Static Hyper Prior Decoding Quantization}

The hyper prior decoder takes the encoded hyper prior and generates the mean and scale of the probability distributions for the main latents. By quantizing the hyper prior decoding path, arithmetic operations are restricted to 8-bit integer-based representations, making the process more deterministic and less susceptible to precision mismatches across different platforms. This is crucial to ensure that both the encoder and decoder generate consistent mean and scale parameters for entropy encoding/decoding the main latents, as any discrepancies in these parameters could lead to significant errors in reconstruction, as shown in Fig.~\ref{fig_fp_error}. 

%Therefore, quantizing the hyper prior decoding path mitigates floating-point precision mismatches, ensuring that both the encoder and decoder interpret the probability distributions of the main latents consistently.

%, resulting in reliable and accurate reconstruction across different platforms.

\subsection{Quantization Parameters Exhaustive Search}

As previously mentioned in Section~\ref{sec_pytorch}, PyTorch stipulates many possible configurations for the quantization process. This is due to the variety of observers and quantization schemes available. In order to determine the optimal configuration which yields the best results in terms of compression efficiency, an exhaustive search was conducted through different quantization settings for the SSF hyper prior decoding path. Altogether, 400 different configurations were considered, combining various observers and quantization schemes for both network weights and activations. It is important to mention that some configurations could not be evaluated due to incompatible combinations of observers and quantization schemes.

For each quantization configuration, the model was calibrated using a random subset of 78 sequences from the Vimeo90k~\cite{Vimeo90k}, converted to a quantized version, and then tested using a different random subset of 78 sequences from the same dataset. In both phases, the videos were cropped to $256\times256$ pixels. During the calibration process, a random cropping strategy was applied, while for testing, the videos were always center-cropped. It is important to note that this process was repeated for each quality level, as they employ different network weights, meaning each configuration was evaluated nine times to generate nine points on the rate-distortion curve.

\subsection{Exhaustive Search Results} \label{sec_esr}

\begin{figure}[tbp]
    \centerline{\includegraphics[width=.45\textwidth]{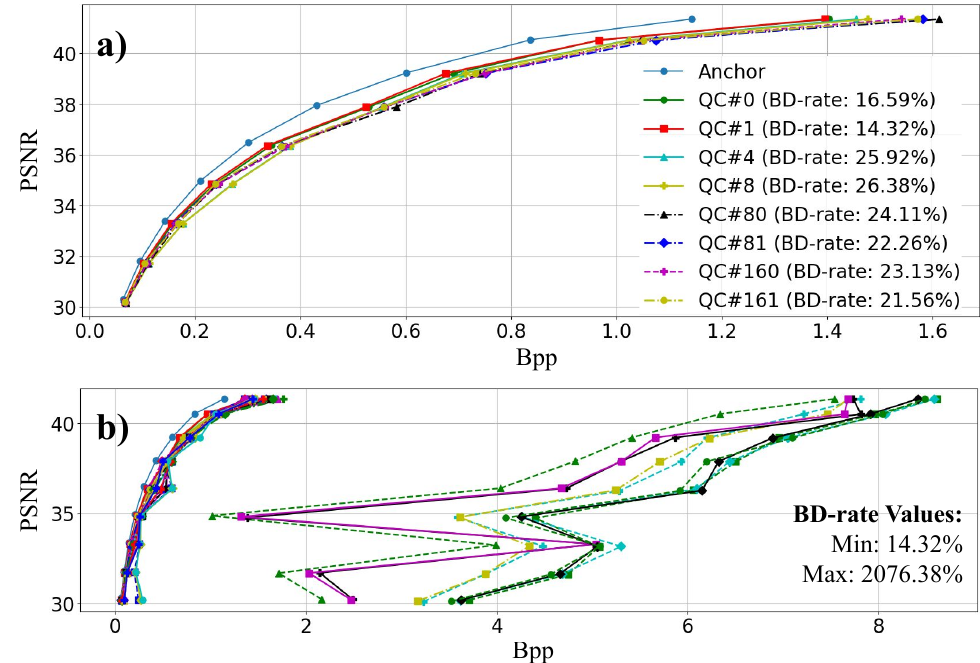}}
    \caption{Rate-distortion curves for: a) all quantization configurations; b) configurations with a BD-rate increase of less than 30\%.}
    \label{fig_result_analysis}
\end{figure}

Each configuration was compared with the original SSF codec (without quantization) in terms of Bjøntegaard Delta Rate (BD-rate)~\cite{BDrate}, where lower BD-rate values indicate better compression efficiency. Out of the 400 evaluated configurations, most encountered compatibility issues. In some cases, a given quantization scheme could not be paired with a particular observer, while in other cases, there were incompatibilities between certain observers and the transpose convolution modules used in the hyper prior decoder. Ultimately, 36 configurations produced results, which are shown in Fig.~\ref{fig_result_analysis}(a). Some configurations resulted in significantly high BD-rate values, indicating that the quantization scheme applied to the hyper prior decoding path can severely affect the compression efficiency. For readability purposes, Fig.~\ref{fig_result_analysis}(b) highlights the configurations with a BD-rate increase of less than $30\%$. One may notice that Quantization Configuration (QC) \#1 achieved the best results in terms of compression efficiency decrease ($14.32\%$) in comparison with all the others. 

Table~\ref{table_conf} presents the quantization configurations used in the top four lowest and highest BD-rate increase scenarios. In the table, ``QS" stands for quantization scheme, ``PTS" for \textit{Per Tensor Symmetric}, ``PTA" for \textit{Per Tensor Affine}, and ``MA MinMax" refers to the \textit{Moving Average MinMax} observer. It is noteworthy that all best results were achieved using a PTS quantization scheme for activations, while the worst results were obtained with a PTA (asymmetric) scheme. This highlights that the activations in the hyper prior decoding path should be quantized using a symmetric quantization scheme. Additionally, as shown in Table~\ref{table_conf}, the two best results employed the \textit{Histogram} observer for both activations and weights, whereas none of the worst results used the \textit{Histogram} observer for weights. This indicates that, for the hyper prior decoding path, the weights should be quantized using the \textit{Histogram} observer for optimal performance.

% The next section presents experimental results and an comprehensive analysis of the SSF quantized model using configuration 1, which yielded the best results in terms of BD-rate increase.

\begin{table}[tbp]
\caption{Quantization Observers and Schemes}
\label{table_conf}
\scriptsize{
\begin{center}
\begin{tabular}{c|c|c|c|c|c}
\hline
\hline
\textbf{Quant.} & \multicolumn{2}{|c|}{\textbf{Activations}} & \multicolumn{2}{|c|}{\textbf{Weights}} & \textbf{BD} \\
\cline{2-5} 
\textbf{Config.} & \textbf{\textit{Observer}}& \textbf{\textit{QS}}& \textbf{\textit{Observer}}& \textbf{\textit{QS}} & \textbf{Rate} \\
\hline
\hline
1 & \textit{Histogram} & PTS & \textit{Histogram} & PTA & 14.3\% \\
\hline
0 & \textit{Histogram} & PTS & \textit{Histogram} & PTS & 16.6\% \\
\hline
161  & \textit{MA MinMax} & PTS & \textit{Histogram} & PTA & 21.6\% \\
\hline
81  & \textit{MinMax} & PTS & \textit{Histogram} & PTA & 22.3\% \\
% \hline
% 160  & \textit{MA MinMax} & PTS & \textit{Histogram} & PTS & 23.1\% \\
\hline
\hline
%  25 & \textit{Histogram} & PTA & \textit{MinMax} & PTA & 1794\% \\
% \hline
189  & \textit{MA MinMax} & PTA & \textit{MA MinMax} & PTA & 2021\% \\
\hline
185  & \textit{MA MinMax} & PTA & \textit{MinMax} & PTA & 2045\% \\
\hline
109  & \textit{MinMax} & PTA & \textit{MA MinMax} & PTA & 2074\% \\
\hline
105 & \textit{MinMax} & PTA & \textit{MinMax} & PTA & 2076\% \\
\hline
\hline
\end{tabular}
\label{tab1}
\end{center}}
\end{table}

\section{Experimental Results}

The results evaluated in this section employs the SSF codec with a 8-bit integer quantized hyper prior decoding path. The quantized codec was used for both generating and decoding the encoded bitstream. A total of 32 frames at $1920\times1080$ resolution from each sequence of UVG and HEVC-B datasets~\cite{HEVC-B} were encoded. 

\begin{figure}[tbp]
\centerline{\includegraphics[width=.45\textwidth]{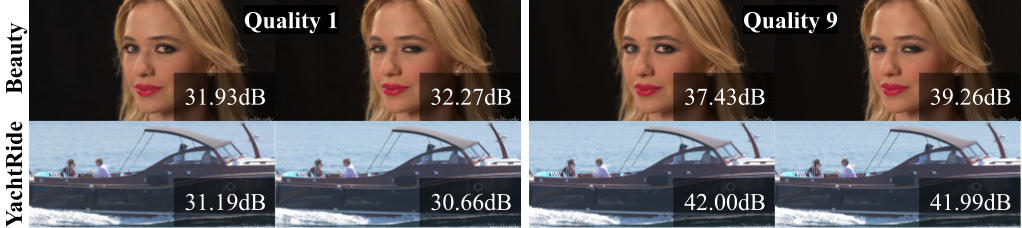}}
\caption{Mitigation of floating-point error propagation across different SSF quality levels due to hyper prior decoding path quantization.}
\label{fig_fp_non_error}
\end{figure}

Fig.~\ref{fig_fp_non_error} shows the reconstructed frames for the same scenario as in Fig.~\ref{fig_fp_error}, but with the quantized hyper prior decoding path. The image is no longer catastrophically degraded, confirming the effectiveness of the proposed solution.

\begin{figure}[tbp]
\centerline{\includegraphics[width=.48\textwidth]{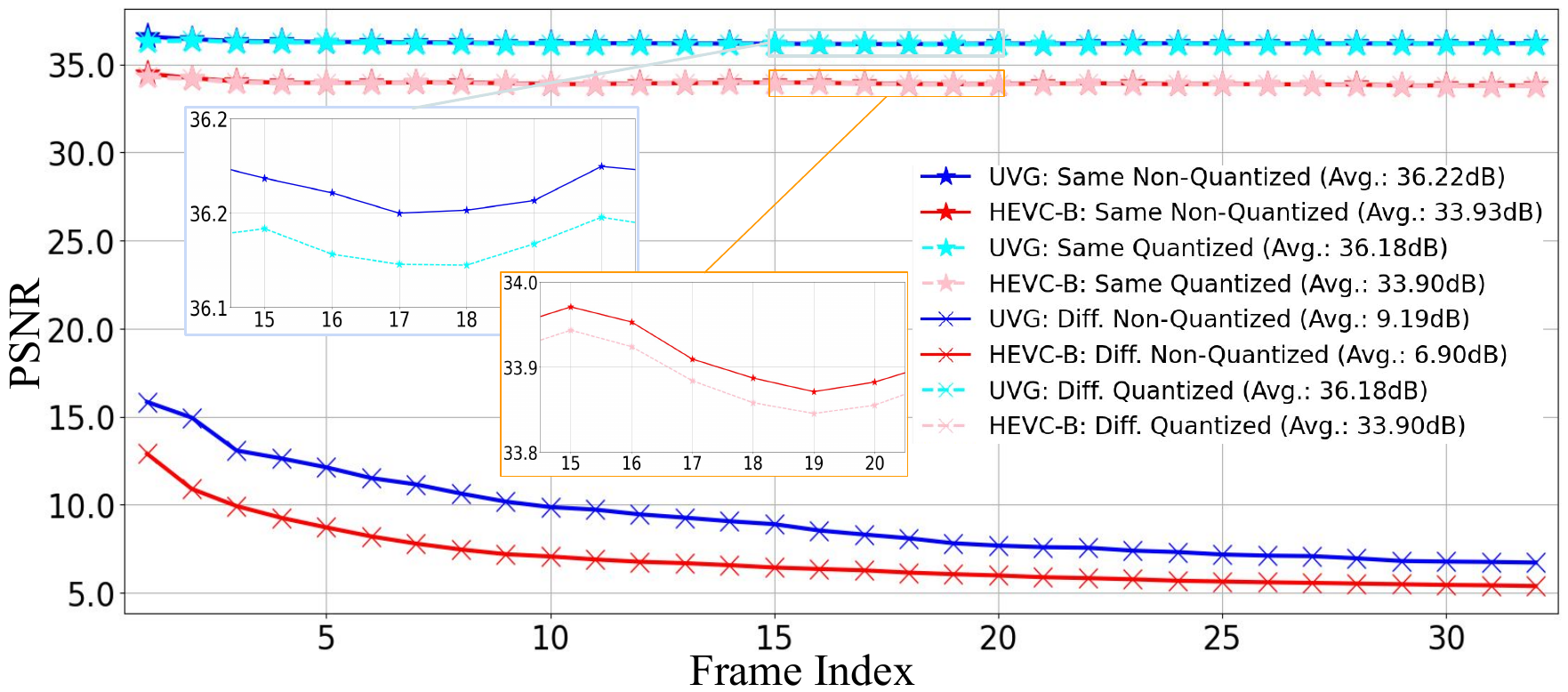}}
\caption{PSNR curves for four scenarios comparing quantized and non-quantized SSF versions when decoding the bitstream generated on same and different machines.}
\label{fig_psnr_frameindex}
\end{figure}

Fig.~\ref{fig_psnr_frameindex} presents the PSNR curves of reconstructed frames for eight distinct scenarios, considering whether the encoding and decoding processes are performed on the same (\textit{Same}) or different (\textit{Diff}) machines, and comparing the SSF versions (\textit{Quantized} or \textit{Non-Quantized}). This experiment was conducted using quality level 5 of the model and both UVG and HEVC-B dataset sequences. As observed, without quantization, the PSNR significantly drops when the sequence is decoded on a different machine. Additionally, the PSNR decrease caused by quantization is negligible ($\leq0.04$dB) and is more noticeable in the first frame, which is intra encoded. 
Finally, there is a negligible difference in the reconstructed frames when using the quantized model to decode the bitstream on the same or different machines ($<0.001$dB). This is due to the fact that the VAE path was not quantized, which can result in minor variations in the image reconstruction. 
Although one may expect that this mismatch error in the VAE decoding path may drift and accumulate along the temporal dimension via inter-frame prediction, this effect is not as considerable as the hyper prior decoding path in the case of SSF. Nonetheless, in cross-platform scenarios, the quantized version improves the reconstructed image quality by about 27dB compared to the non-quantized version.

Regarding compression efficiency for the nine quality points, the proposed solution presents a slight BD-PSNR decrease of 0.2913dB / 0.2776dB, and a BD-rate increase of 9.93\% / 9.02\% for the UVG / HEVC-B datasets. Note that BD-rate results differ to the one reported in Section~\ref{sec_esr}. It is due to fact that the earlier analysis was conducted on a distinct dataset with lower spatial resolutions and fewer frames. This impact on compression efficiency occurs considering the encoding and decoding processes are performed on the same machine. When using different machines, the non-quantized codec results in significantly degraded image quality, making BD-rate and BD-PSNR evaluation impractical. Therefore, in real scenarios, the proposed solution also presents a compression efficiency enhancement.

%Another important aspect to consider is a further hardware design of the SSF decoder. Switching the hyper prior decoding path operations from 32-bit floating-point numbers to 8-bit integers reduces the area size and static power dissipation, as floating-point operators require more complex architectures to handle the sign, exponent, and mantissa. Additionally, 8-bit integer operators rely on simpler logic gates, resulting in fewer switching activities, which decreases dynamic power dissipation, leading to a total power dissipation reduction. Therefore, by quantizing the hyper prior decoding path, it enables the bitstream to be decoded on different platforms, while reducing the SSF decoder power dissipation.

\section{Concluding Remarks}
In this paper, we first demonstrated that current learning-based video codecs, specifically the SSF codec, are not yet suitable for real-world applications due to floating-point round-off errors. By applying static quantization to the hyper prior decoding path, we showed that this issue can be mitigated with negligible impact on video compression efficiency. In addition to making the codec more practical for real-world use, replacing floating-point operations with integer ones could offer the beneficial side effect of reducing power dissipation in future hardware designs. For future work, we plan to extend this study to context- and transformer-based codecs.

\end{document}